# Tuning electronic heat transport in graphene/metal heterostructures with ultralow thermal conductivity

Bin Huang[1,2], Weidong Zheng[1,2], Yee Kan Koh[1,2]*

[1]*Department of Mechanical Engineering, National University of Singapore, Singapore 117576*

[2]*Centre for Advanced 2D Materials, National University of Singapore, Singapore 117542*

*Correspondence to: mpekyk@nus.edu.sg

# Abstract

Prior ultralow thermal conductivity ($\Lambda$) materials(*1-7*) are not suitable for thermoelectric applications due to the limited electronic transport in the materials. Here, we present a new class of ultralow $\Lambda$ materials with substantial electronic heat transport. Our samples are graphene/metal heterostructures of transferred graphene and ultrathin metal films (Pd, Au and Ni) deposited by either thermal evaporation or rf magnetron sputtering. For the evaporated samples, we achieve an ultralow $\Lambda$ of 0.06 W m$^{-1}$ K$^{-1}$, with phonons as the dominant heat carriers. The ultralow $\Lambda$ is due to a huge disparity(*8, 9*) in phonon energy in graphene and metals. Interestingly, for the sputtered samples, we find that $\approx$50 % of heat is carried by electrons, even when $\Lambda \approx$ 0.1 W m$^{-1}$ K$^{-1}$. We attribute the electronic contribution to transmission of electrons across atomic-scale pinholes(*9*) in graphene. With the ultralow $\Lambda$ and substantial electronic transport, the new materials could be explored for thermoelectric applications.

# Text

Traditionally, the lowest thermal conductivity is found in amorphous materials in which heat is carried by quantized vibration modes that randomly diffuse and relax with a relaxation rate on the order of their vibration frequencies(*10*). Over the past decade, novel layered structures with ultralow thermal conductivity have been proposed and demonstrated to beat the amorphous limit, mainly due to scattering of phonons by a high density of interfaces(*1, 2, 4, 6*). For these layered structures, the thermal conductivity can be approximated by a simple expression assuming that the thermal resistance of interfaces adds in series; $\Lambda = Gd/2$, where $G$ is the thermal conductance of a single interface and $d$ is the period of the layered structures. Thus, ultralow thermal conductivity, in principle, could be achieved by choosing interfaces that effectively impede transmission of phonons (low $G$) and using thin films (low $d$). However, when $d$ is surpassingly small (e.g., ≈3 nm), this simple expression breaks down and the thermal conductivity saturates(*2, 11*) because low-energy phonons can transmit across the interfaces coherently or incoherently without being scattered(*11-13*). As a result of the transmission of low-energy phonons, interfaces are no longer effective to reduce the thermal conductivity(*11*) and the thermal conductivity of short-period superlattices usually plateaus at a value much higher than that given by $Gd/2$(*2*). In this regard, graphene is an ideal candidate to achieve ultralow thermal conductivity, since graphene is only one atomic layer thick and is effective in blocking transmission of low-energy phonons due to the lack of low-energy modes in graphene(*8, 14*). Here, we successfully demonstrate the use of graphene for thermal insulation through a series of fully dense graphene/metal heterostructures with a thermal conductivity and a thermal diffusivity as small as 0.06 W m$^{-1}$ K$^{-1}$ (≈2 times of that of the air) and 2.6×10$^{-4}$ cm$^2$ s$^{-1}$ respectively, comparable to the world record lowest values.

Despite the low thermal conductivity, prior ultralow thermal conductivity materials are not suitable for thermoelectric applications. To qualify to be efficient thermoelectric materials, two criteria in term of heat transport (in addition to the requirement for a high Seebeck coefficient) must be met, i.e., a low thermal conductivity and a high electronic contribution. The criteria can be understood by evaluating a figure-of-merit for thermoelectric efficiency, *ZT*. Often, *ZT* is expressed as $ZT = S^2\sigma T/\Lambda$, where *S* is Seebeck coefficient, *T* is temperature, and $\sigma$ and $\Lambda$ are electrical and thermal conductivity, respectively. For a power factor of $S^2\sigma \approx 25$ μW cm$^{-1}$ K$^{-2}$ (a typical value for good thermoelectric materials) and $ZT \approx 1$ at room temperature, this translates to $\Lambda \approx 1$ W m$^{-1}$ K$^{-1}$. Thus, the criterion of a low $\Lambda$ can be easily met by the ultralow thermal conductivity materials. However, *ZT* can also be expressed as $ZT = (S^2/L)/(1 + \Lambda_{ph}/\Lambda_e)$, where $L = 2.44\times10^{-8}$ W Ω K$^{-2}$ is the Lorenz number, and $\Lambda_{ph}$ and $\Lambda_e$ are the phononic (lattice) and electronic components of the thermal conductivity, respectively and $\Lambda = \Lambda_{ph} + \Lambda_e$. In other words, *ZT* approaches a limit of $S^2/L$, but is reduced by a large ratio of phononic to electronic thermal conductivity. In good thermoelectric materials, the reduction due to $\Lambda_{ph}/\Lambda_e$ is minimized and thus electronic heat transport is always substantial. In this respect, we successfully prepared graphene/metal heterostructures with $\Lambda_{ph}/\Lambda_e$ of only $\approx 1$, despite the ultralow thermal conductivity of $\approx 0.1$ W m$^{-1}$ K$^{-1}$. We attribute the high electronic transport to transmission of electrons across atomic-scale pinholes in graphene. Since Seebeck coefficient across atomic-scale pinholes(*15, 16*) and Au/graphene interfaces(*17*) could be substantial, our graphene/metal heterostructures could provide a new route to achieve efficient thermoelectric materials.

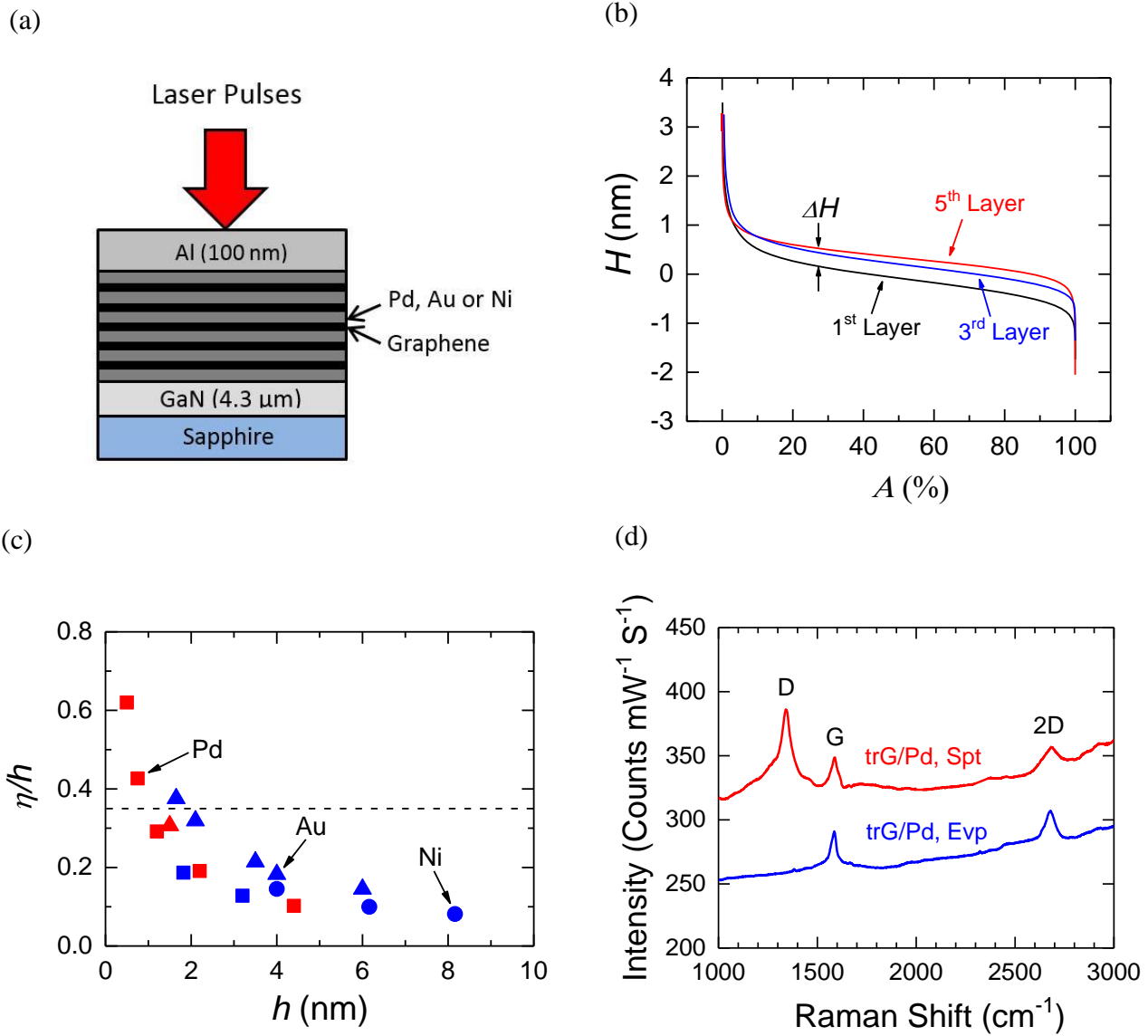

**Fig. 1. (a)** A schematic diagram showing the cross sectional view of our metal/graphene heterostructures. The heterostructures consist of 5 layers of CVD-grown transferred graphene (trG) and 6 layers of metals (i.e., Pd, Au or Ni) stacked alternatingly with one another. **(b)** Determination of the conformity of graphene in the Pd/metal ($h$ = 1.3 nm) heterostructures by rf magnetron sputtering, from plots of relative height $H$ as function of accumulative percentage area $A$. Transferred graphene is considered conformal when $\Delta H$ compared to previous layer is < 0.5 nm, see ref. 19 for more discussion. **(c)** Ratio of root mean squared (RMS) roughness and thickness as a function of thickness for Au (triangles), Pd (squares) and Ni (circles) metal films on thermal oxide substrates, deposited either by thermal evaporation (blue) or by rf magnetron sputtering (red). This series of samples were separately prepared to determine the thinnest continuous metal films that we can deposit. The dashed line corresponds to $\eta/h$ = 0.35, which is the ratio that we assume the film to be non-continuous. **(d)** Raman spectra of Pd/trG

heterostructures prepared by thermal evaporation (Evp, $h = 1.8$ nm) and rf magnetron sputtering (Spt, $h = 1.3$ nm), as labeled.

Our graphene/metal heterostructures consist of alternating layers of monolayer transferred graphene (trG) and ultrathin metal films (Pd, Au and Ni), see Fig. 1(a). To build the heterostructures, we transfer monolayer graphene onto each thin metal film (deposited by thermal evaporation or rf magnetron sputtering, see below), and repeat the processes until the heterostructures are sufficiently thick for the thermal measurements. Our graphene was purchased from Graphene Supermarket and was grown by chemical vapor deposition (CVD). We perform graphene transfers with poly(bisphenol A carbonate) (PC) as the support layers(*18, 19*), see ref. 19 for the procedures of our graphene transfer. We ensure that our transfers are clean by checking the topographic images acquired by atomic force microscope (AFM). We confirm that the amounts of PC residues and wrinkles on the transferred graphene are reasonably small even after 5 graphene transfers, see the AFM topographic images in Fig. S1 in Supplementary Information. We also quantify conformity of the transferred graphene to the underlying metal films, using an approach that we developed, see ref. 19 for the details. In the approach, we derive the accumulative percentage of areas ($A(H)$) from the highest peaks to a certain height from the depth histograms of AFM topography images, and compare the relative height $H$ vs $A(H)$ of graphene after each transfer in Fig. 1(b). We find that the height profiles are very similar even after multiple graphene transfers and metal deposition, see Fig. 1(b), and $\Delta H$ for graphene after the first and fifth transfer is consistently <0.6 nm even when $A(H) \approx 100$ %. We thus conclude that our transferred graphene is $\approx 100$ % conformal.

We deposit the ultrathin metal films in the graphene/metal heterostructures by either thermal evaporation or rf magnetron sputtering. To determine the average thickness of the films in the heterostructures, we measure the total thickness of metal films on thermal oxide substrates that were included in the chamber during the deposition, see Methods in the Supplementary Information for details. To achieve the thinnest metal films and thus the lowest thermal conductivity, we deposit a series of ultrathin metal films with different intended thicknesses on thermal silicon dioxide, and measure the root-mean-square (rms) roughness $\eta$ of the thin films by AFM. We then plot the ratios of the rms roughness to film thickness $h$, $\eta/h$, of the films in Fig. 1(c). We approximate from the rms of a perfect sine curve that the metal films are continuous when $\eta/h < 0.35$. We thus estimate that the thinnest continuous metal films for Pd and Au that we can achieve are roughly $\approx 1$ nm and $\approx 2$ nm, respectively. These estimated values are consistent with our measurements of the electrical resistivity of the films by van der Pauw method, see Fig. S4. The electrical resistivity increases substantially (e.g., $> 100$ $\mu\Omega$-cm) when the deposited metals are just disconnected islands. We thus find that this corresponding to a thickness of $\approx 1$ nm for Pd and $\approx 2$ nm for Au, similar to what we derived from Fig. 1c. We emphasize that the rough estimations of the thinnest continuous films we can deposit are only meant to guide us on the periods of the trG/metal heterostructures that we should prepare; our thermal measurements of the heterostructures confirm that the ultrathin metal films have sufficient coverage to thermally impede the heat flow, see discussion below.

To check the quality of the transferred graphene (trG) in our trG/metal heterostructures, we performed Raman spectroscopy on two trG/Pd heterostructures (with the Pd films either evaporated or sputtered), see Fig. 1(d). We find that while the graphene in the evaporated sample remains pristine with no significant D peak in its Raman spectrum, the graphene in the sputtered sample was however damaged with a huge D peak in its Raman spectrum; see Fig. 1(d). The results are similar to what we obtained in our previous Raman measurements on metal/graphene/SiO$_2$ interfaces(*9*). In our previous studies(*9*), we also performed high resolution

AFM using a sharp tip with a nominal radius of 5 nm and a force constant of 17 N m$^{-1}$ on an ion-bombarded graphene sample that has a similar Raman spectrum as the sputtered graphene here. We found that the phase-contrast images of the ion-bombarded graphene and pristine graphene are almost identical, indicating that the damages that ion-bombardment creates are on atomic scale. We thus infer that the only difference between our sputtered and evaporated trG/metal heterostructures is that sputtering creates atomic-scale pinholes in graphene.

We measure the thermal conductivity ($\Lambda$) of the trG/metal heterostructures by time-domain thermoreflectance (TDTR), see Methods in the Supplementary for details. In our measurements, we employ a pump-leak correction approach to eliminate artifact signals from the diffusely scattered pump beam[20]. We plot $\Lambda$ of the trG/Pd, trG/Au and trG/Ni heterostructures in Fig. 2(a), as a function of the period ($d$) of the heterostructures (i.e., sum of the thickness of a single layer of metal film and graphene). In the same figure, we also plot the thermal conductivity of disordered WSe$_2$[1], self-assembled organoclay nanolaminates[6], hybrid organic-inorganic zincones[5], and W/Al$_2$O$_3$ nanolaminates[2] for comparison. We find that the lowest $\Lambda$ of our heterostructures is $\approx$0.06 W m$^{-1}$ K$^{-1}$ for a heterostructure of graphene and 2.1 nm thick evaporated Au.

To understand $\Lambda$ of the heterostructures, we compare our measurements to a simple estimation of $\Lambda = Gd/2$ in Fig. 2(a). Here, $G$ is not a fitting parameter, but measured thermal conductance of the corresponding metal/trG/metal interfaces[9, 14]. We find that, in contrast to W/Al$_2$O$_3$ nanolaminates[2], our measurements agree well with the simple estimation of $\Lambda = Gd/2$, even when $d = 2$ nm, and no plateau of $\Lambda$ at short-period heterostructures is observed, see Fig. 2(a). The linear dependence confirms that graphene is effective to impede heat flow by low-energy phonons even in a heterostructure, unlike other superlattices and multilayers that low-energy phonons transmit without being strongly scattered[11-13]. We propose that the effectiveness of graphene to block low-energy phonons could be due to scarcity of low-energy phonon modes in graphene; since graphene interfaces are essentially decoupled(*8, 14*) and no tunneling of phonons were previously reported, propagation of phonons in the trG/metal heterostructures is limited by the lowest number of low-energy modes that are available in the heterostructures (i.e., in graphene).

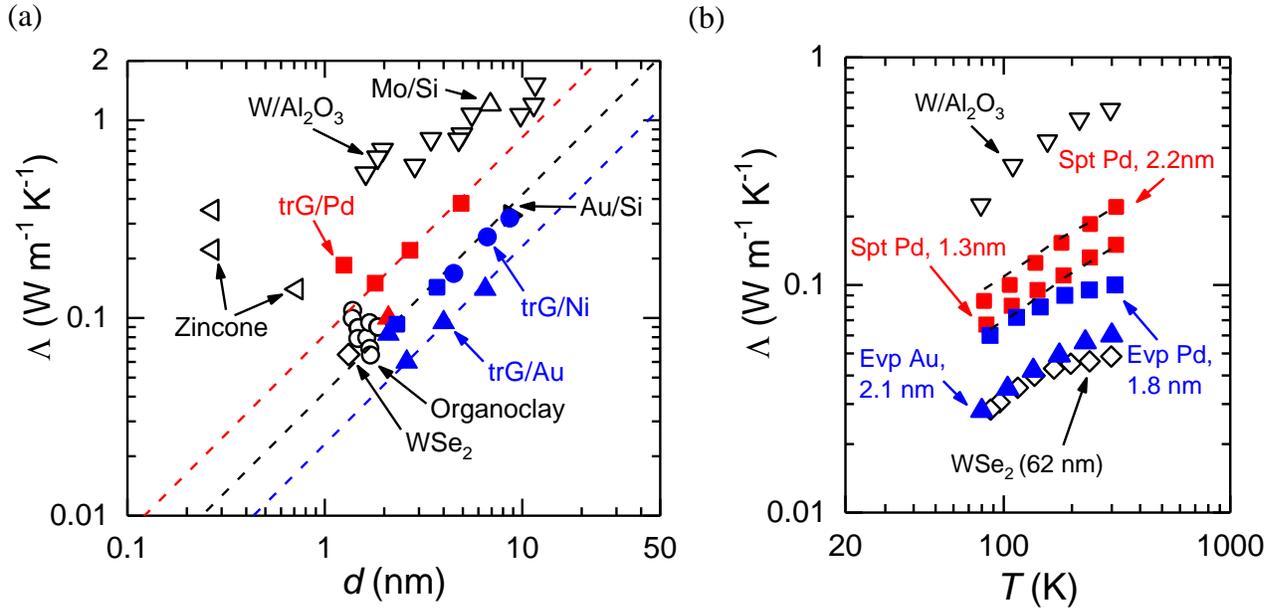

**Fig. 2. (a)** Thermal conductivity Λ of Pd/trG (squares), Au/trG (triangles), and Ni/trG (circles) heterojunctions with the metal layers deposited by either thermal evaporation (blue) or rf magnetron sputtering (red), plotted against the period *d* of the structure. The thermal conductivity of W/Al$_2$O$_3$ nanolaminate (open down triangles, ref. 2), Organoclay nanolaminate (open circles, ref. 6), Zincone thin films (open left triangles, ref. 5), Au/Si multilayer (open right triangle, ref. 3), Mo/Si multilayer (open up triangle, ref. 4) and disordered WSe$_2$ (open diamond, ref. 1) are also plotted for comparison. The dashed lines are calculations of equation, $\Lambda = Gd/2$, using $G$ = 82, 42, 23 MW m$^{-2}$ K$^{-1}$, which are the thermal conductance of sputtered Pd/trG/Pd, evaporated Pd/trG/Pd and evaporated Au/trG/Au interfaces, respectively. **(b)** Temperature dependence of the thermal conductivity Λ of heterostructures of graphene and Pd or Au thin films, prepared by either rf magnetron sputtering (labeled as "Spt") or thermal evaporation (labeled as "Evp"). The thickness of the metal films (*h*) are as labeled. The thermal conductivity of W/Al$_2$O$_3$ nanolaminate (open squares, ref. 2) and disordered WSe$_2$ (open diamond, ref. 1) are also plotted for comparison. The dashed lines are fitting of the measurements assuming that the excess heat is carried by electrons.

We observe two clear outlier data points that do not follow the linear dependence in Fig. 2a, one Pd/trG heterostructure with $d = 1.3$ nm, prepared by magnetron sputtering, and one Au/trG heterostructure with $d = 2.1$ nm, prepared by thermal evaporation. The periods of 1.3 nm and 2.1 nm correspond to metal film thicknesses of 0.8 nm and 1.6 nm, respectively. From Fig. 1c, we find that the Pd film prepared by rf magnetron sputtering with $h = 0.8$ nm and Au film prepared by thermal evaporation with $h = 1.6$ nm are not continuous. We thus believe that the higher $\Lambda$ of the short-period samples is not due to transport of coherent phonons (e.g., in other superlattices(*12, 13*)) across the interfaces, but due to enhancement of heat flow by direct contacts of graphene not properly separated by discontinuous metal films.

Interestingly, for a similar period $d$, we find that $\Lambda$ of samples prepared by sputtering is substantially higher than those prepared by thermal evaporation, see Fig. 2(a). To understand the origins of the difference, we measure the temperature dependence of $\Lambda$ of two Pd/trG heterostructures prepared by rf magnetron sputtering and a Pd/trG heterostructure prepared by thermal evaporation, see Fig. 2(b). We find that while heterostructures prepared by thermal evaporation exhibit a weak temperature dependence indicating that heat is carried by phonons, heterostructures prepared by sputtering have a strong temperature dependence suggesting that electrons contribute appreciably. We thus fit $\Lambda$ of the sputtered samples with $\Lambda = \Lambda_{ph} + LTd/\rho_c$, where $\Lambda_{ph}$ and $\Lambda_e = LTd/\rho_c$ are phononic and electronic components of thermal conductivity respectively, and $\rho_c$ is electrical specific contact. We estimate $\Lambda_{ph}$ from $\Lambda$ of the evaporated samples assuming that the thermal conductivity is proportional to the period. We set $\rho_c$ as the only fitting parameter and derive $\rho_c \approx 2.2 \times 10^{-9}$ $\Omega$ cm$^2$, similar to $\rho_c$ we previously derived from fitting of the thermal conductance of Pd/trG/Pd interfaces(*9*). For the trG/Pd heterostructure prepared by rf magnetron sputtering with $d = 1.8$ nm, we thus estimate that both $\Lambda_{ph}$ and $\Lambda_e$ are approximately 0.075 W m$^{-1}$ K$^{-1}$ at room temperature, giving a $\Lambda_e / \Lambda_{ph}$ ratio of $\approx 1$.

Finally, we evaluate the potential of our trG/metal heterostructures as thermal insulators and thermoelectric materials in Fig. 3. For heat insulation under transient conditions, retardation of oscillating heat flow is determined not by the thermal conductivity but by the thermal diffusivity, $\alpha = \Lambda / C_V$ where $C_V$ is volumetric heat capacity. Thus, we compile the thermal conductivity of a wide range of ultralow thermal conductivity materials, including layered structures(*1-3, 5, 6*), fullerenes(*7, 21*), amorphous materials(*22, 23*), and plot it as a function of $C_V$ in Fig. 3(a). We find that $\alpha = 2.6 \times 10^{-4}$ cm$^2$ s$^{-1}$ of our trG/Au heterostructures is comparable to the world record lowest value(*1, 6*), see Fig. 3(a).

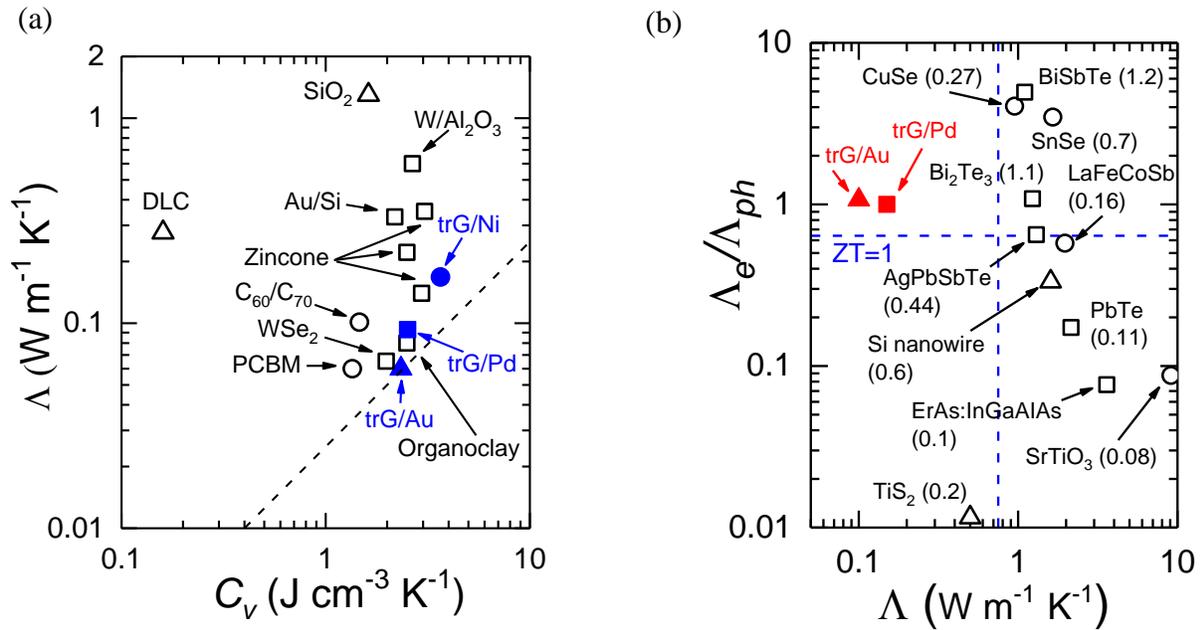

**Fig. 3. (a)** thermal conductivity $\Lambda$ of Pd/trG (solid squares, $h = 1.8$ nm), Au/trG (solid triangle, $h = 2.1$ nm), and Ni/trG (solid circles, $h = 4.0$ nm) heterostructures with the metal layers deposited by thermal evaporation plotted as a function of volumetric heat capacity $C_v$. We compare our measurements to W/Al$_2$O$_3$ nanolaminate (ref. 2), Organoclay nanolaminate (ref. 6), Zincone thin films (ref. 5), Au/Si multilayer (ref. 3), disordered WSe$_2$ (ref. 1), [6,6]-phenyl-C$_{61}$-butyric acid methyl ester (PCBM) (ref. 7), C$_{60}$/C$_{70}$ bucky balls (ref. 21), amorphous silicon dioxide (SiO$_2$) (ref. 22), diamondlike carbon (DLC) (ref. 23). We use open squares, circles and triangles to represent layered structures, fullerene and amorphous material respectively. The dashed line is the calculation of equation $\Lambda = C_v \times \alpha$ using a thermal diffusivity $\alpha$ value of $2.5 \times 10^{-8}$ m$^2$ s$^{-1}$. **(b)** Compilation of ratios of electronic thermal conductivity and lattice thermal conductivity $\Lambda_e/\Lambda_{ph}$ as a function of total thermal conductivity $\Lambda$ of our sputtered Pd/trG (solid squares, $h = 1.3$ nm) and sputtered Au/trG (solid triangle, $h = 1.6$ nm) heterostructure samples at 300 K. We compare our results to thermoelectric (TE) measurements obtained from literature (open symbols). Open circles, triangles and squares represent bulk TE materials,[24-27] nanostructures TE materials[33,34] and bulk TE materials with nanostructures precipitates,[28-32] respectively. The figures in the bracket are the dimensionless figure of merit (ZT) values of the respective thermoelectric materials. The blue lines are values of $\Lambda$ and $\Lambda_e / \Lambda_{ph}$ that give a $ZT$ of $\approx 1$, see the text for discussion.

For thermoelectric applications, two important criteria are a low $\Lambda$ and a high $\Lambda_e / \Lambda_{ph}$, as explained in the introduction of the paper. We thus plot the $\Lambda_e / \Lambda_{ph}$ vs $\Lambda$ for our sputtered trG/metal heterostructures, compared to a wide range of thermoelectric materials (bulk(*24-27*), bulk with nanoscale precipitates(*28-32*), nanostructures(*33, 34*)) in Fig. 3(b). In the plot, we also include approximated values of $\Lambda$ and a high $\Lambda_e / \Lambda_{ph}$ to achieve a $ZT$ of $\approx 1$, see the blue dashed lines in the plot. (We estimate $\Lambda$ to achieve $ZT \approx 1$ using equation $ZT = S^2 \sigma T / \Lambda$ and an assumed $S^2 \sigma \approx 25$ µW cm$^{-1}$ K$^{-2}$. We estimate $\Lambda_e / \Lambda_{ph}$ to achieve $ZT \approx 1$ using equation $ZT = (S^2/L)/(1 + \Lambda_{ph}/\Lambda_e)$ and an assumed Seebeck coefficient of 250 µV K$^{-1}$). As suggested in Fig. 3(b), both criteria are met by our trG/metal heterostructures, and thus the novel structure could be explored for thermoelectric applications.

In summary, we present novel graphene/metal heterostructures with an ultralow thermal conductivity of $\approx 0.1$ W m$^{-1}$ K$^{-1}$ and a substantial electronic heat transport of $\Lambda_e / \Lambda_{ph} \approx 1$. We attribute the electronic heat transport to transmission of electrons through atomic scale pinholes in graphene. Due to the low thermal conductivity and a substantial electronic transport, our graphene/metal heterostructures could be explored for thermoelectric applications. Certainly, a high Seebeck coefficient must be achieved for any thermoelectric applications, which is a great challenge for metals. We however think that the Seebeck coefficient could be significantly enhanced at atomic scale, as previously theoretically discussed(*35*) and experimentally demonstrated(*15, 16*). Also, similar heterostructures of graphene and other traditional thermoelectric materials could be fabricated to achieve a high Seebeck coefficient, while taking advantage of the ultralow thermal conductivity exhibited in these heterostructures.

## Acknowledgements

This work is supported by NUS Young Investigator Award 2011 and the Singapore Ministry of Education Academic Research Fund Tier 2, under Award No. MOE2013-T2-2-147. Sample characterization was carried out in part in the Centre for Advanced 2D Materials.

Supplementary Materials for

# Tuning electronic heat transport in metal/graphene heterostructures with ultralow thermal conductivity

Bin Huang[1,2], Weidong Zheng[1,2], Yee Kan Koh[1,2*]

[1]*Department of Mechanical Engineering, National University of Singapore, Singapore 117576*

[2]*Centre for Advanced 2D Materials, National University of Singapore, Singapore 117542*

*Correspondence to: mpekyk@nus.edu.sg

# Methods

## Sample Preparation

Our graphene/metal heterostructures consist of alternating monolayer graphene and ultrathin metal films (Pd, Au and Ni). We bought the chemical vapour deposition (CVD) grown graphene on copper foil with the back side etched from Graphene Supermarket. We use poly(bisphenol A carbonate) (PC) with a molecular weight of 45000 to prepare a 1 wt. % PC in chloroform solution, which is spin-coated onto graphene as the support layer for the transfer. We select PC instead of poly(methyl methacrylate) PMMA due to its smaller molecule size and weaker interfacial adsorption, allowing it to dissolve easier. After spin coating, we float the graphene samples on a 7 wt. % ammonium persulfate (APS) solution to etch away the underside copper. Subsequently, the graphene is cleaned by floating on de-ionized (DI) water. We then scoop the graphene using the metal thin film substrates and soak the samples in chloroform for 24 hours to get rid of the PC layer. Finally, we rinse the samples in isopropanol alcohol (IPA) and blow dry them using dry nitrogen. We repeat this transfer process until the graphene/metal heterostructures are sufficiently thick (e.g., 5 layers of transferred graphene and 6 layers of metal thin films).

Our ultrathin metal films are deposited by thermal evaporation or rf magnetron sputtering. The base pressure of the thermal evaporation and sputtering chamber is $10^{-8}$ and $10^{-7}$ Torr respectively. We used a power density of 1.32 W cm$^{-2}$ at an Argon pressure of 3 mTorr, with a deposition rate of 0.8 Å/s to sputter the ultrathin metal films. For the thermal evaporation, we keep a deposition rate of 0.5 Å/s.

**Thickness Determination**

To determine the thickness of metal films in the graphene/metal heterostructures, we always include a SiO$_2$ (100 nm)/Si substrate with rectangular trenches in the sputtering or evaporation chamber during the preparation of the heterostructures. After all metal films were deposited, the metal films on the SiO$_2$ (100 nm)/Si substrate have a total thickness of the sum of the thicknesses of all metal films in the heterostructure. We then derive the average thickness of metal films in the heterostructures from measurements the thickness of metal films on the substrates.

We accurately measure the thickness of the metal films on the substrates by AFM. Rectangular trenches of size 100 µm x 50 µm are fabricated on SiO$_2$ (100 nm)/Si substrates using the following photolithography method – (a) A layer of positive photoresist S1805 is spincoated on SiO$_2$ substrate and baked dry. (b) The photoresist with an area of 100 µm x 50 µm is exposed to laser with wavelength of 532 nm during laser writing. (c) The trench is formed after the photoresist is been removed during the developing process. After the subsequent deposition of metals (i.e. Pd, Au or Ni) by either thermal evaporation of rf magnetron sputtering, we performed lift off process to remove the photoresist and measure the thickness of the metals using tapping mode AFM.

**Time-domain Thermoreflectance Measurements**

The thermal conductivity of the metal/graphene heterostructures is measured using time-domain thermoreflectance (TDTR) *(20, 36)*. We employed a high frequency of 9.8 MHz, laser 1/e$^2$ radii of 6 µm and a total laser power of 60-90 mW to limit the steady-state temperature rise to <10 K. We measure the total thickness of the metal (i.e., Au, Pd, and Ni) films by tapping mode AFM as mentioned in Thickness Determination and we assume the thickness of each layer of graphene is 0.5 nm *(37)*. The thermal conductivity of the Al film is determined from the electrical resistivity, measured by four point probe, using Wiedemann-Franz law. The room temperature thermal conductance of Al/Pd, Al/Ni and Al/Au interfaces are measured independently, and the values are 190 MW m$^{-2}$ K$^{-1}$, 120 MW m$^{-2}$ K$^{-1}$ and 170 MW m$^{-2}$ K$^{-1}$ respectively. We estimate the thermal conductance of Al/Pd and Al/Au at low temperature by assuming a linear dependence. For Al/Ni interface, we assume a dependence similar to Al/SiO$_2$ since the nickel film would oxidize after exposing to atmospheric conditions. The thermal conductance of Pd/GaN and Au/GaN are obtained from ref. 9. The *G* values are then used in the thermal model during the fitting of the thermal conductivity of metal/graphene heterostructures.

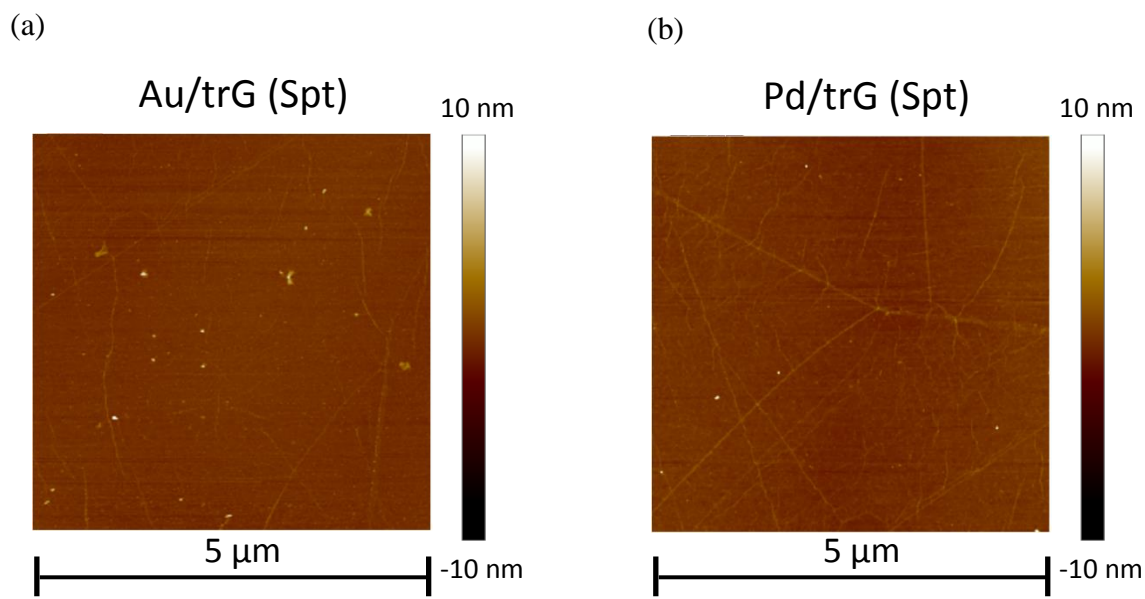

**Fig. S1.** Wet Transfer of Chemical Vapour Deposition (CVD) Grown Graphene. Topographic images of a **(a)** Au/trG ($h$ = 1.6 nm) and **(b)** Pd/trG ($h$ = 1.3 nm) heterostructure fabricated by rf magnetron sputtering (Spt) after the 5$^{th}$ layer of transferred graphene, acquired by tapping mode atomic force microscopy (AFM).

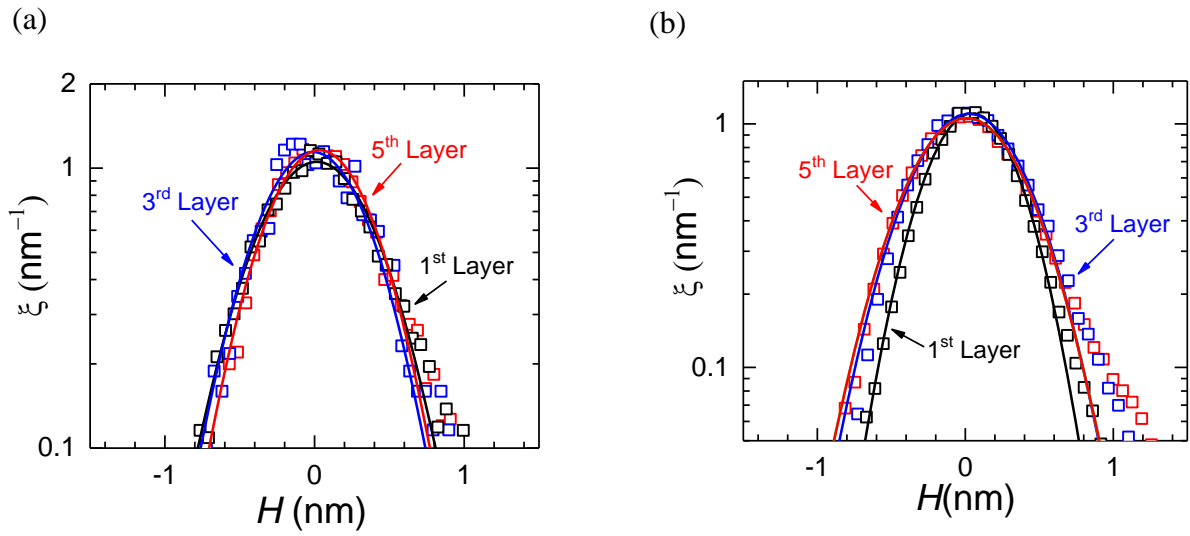

**Fig. S2.** Characterization of Conformity of Graphene. Depth histogram of an **(a)** Au/trG ($h = 1.6$ nm) and **(b)** Pd/trG ($h = 1.3$ nm) heterostructures after the 1st layer, 3rd layer and 5th layer of graphene was transferred on the metal films prepared by rf magnetron sputtering, as labeled. The height distributions are fitted with a Gaussian function (solid lines). We plot the spatial frequency ($\xi$) as a function of relative height $H$; refer to ref. 19 for the definition of spatial frequency ($\xi$).

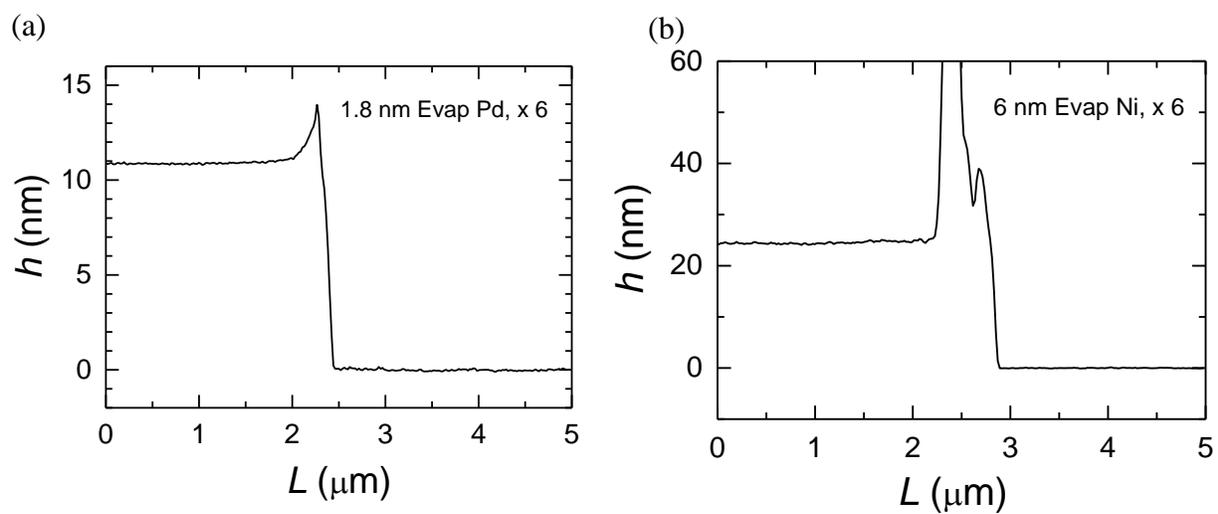

**Fig. S3.** Deposition and measurement of thickness of metal films. Thickness of 6 layers of **(a)** Pd and **(b)** Ni films deposited by thermal evaporation.

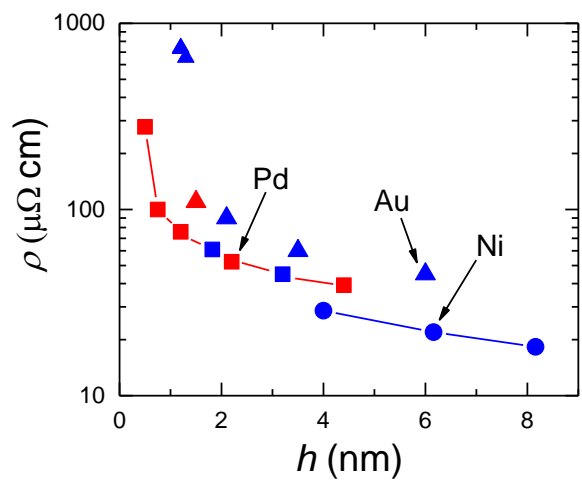

**Fig. S4.** Electrical measurements of ultrathin metal films. Electrical resistivity $\rho$ of ultra-thin Au (triangles), Pd (squares) and Ni (circles) metal films plotted against the film thickness $h$. The metal films are deposited either by thermal evaporation (blue) or by rf magnetron sputtering (red). The metal films are deposited on $SiO_2$ (100 nm)/Si substrates and measured using the standard Van der Paw method to determine the electrical resistivity.

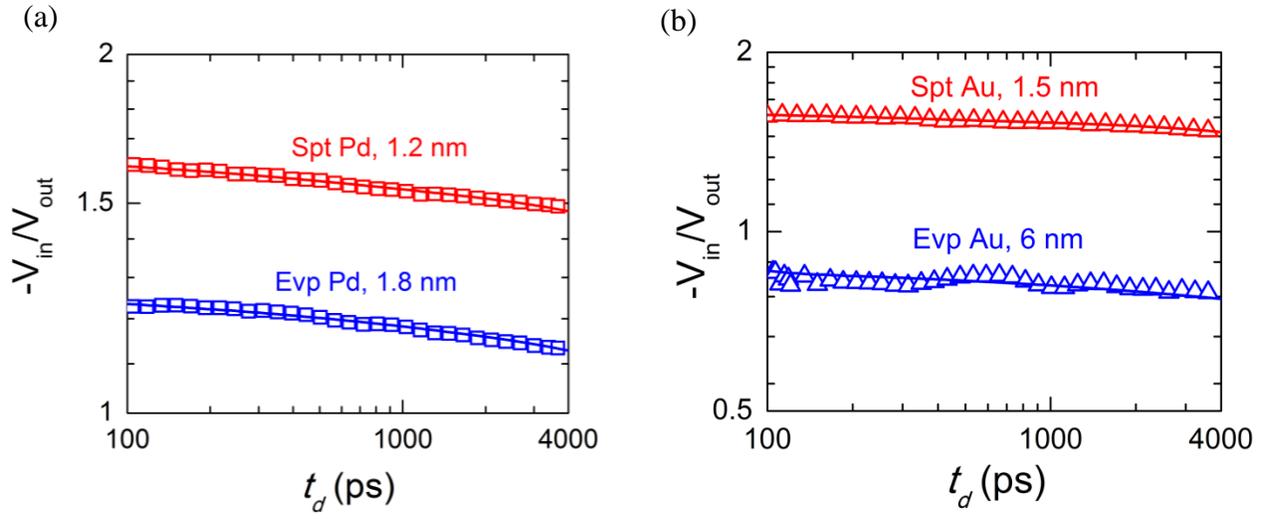

**Fig. S5.** Thermal conductivity measurements of metal/graphene heterostructures by time-domain thermoreflectance (TDTR). Ratios of in-phase and out-of-phase TDTR signals as a function of delay time for **(a)** Pd/trG and **(b)** Au/trG heterostructures fabricated by thermal evaporation (Evp) or rf magnetron sputtering (Spt), as labeled. The solid lines are calculations of a thermal model which is used to fit the measurements.